\documentclass[a4paper,fleqn]{cas-dc}

\usepackage[numbers]{natbib}

\def\tsc#1{\csdef{#1}{\textsc{\lowercase{#1}}\xspace}}
\tsc{WGM}
\tsc{QE}
\tsc{EP}
\tsc{PMS}
\tsc{BEC}
\tsc{DE}

\begin{document}
\let\WriteBookmarks\relax
\def\floatpagepagefraction{1}
\def\textpagefraction{.001}
\shorttitle{Mapping Crisis-Driven Market Dynamics: A Transfer Entropy and Kramers–Moyal Approach to Financial Networks}
\shortauthors{P. Khalilian et~al.}

\title [mode = title]{Mapping Crisis-Driven Market Dynamics: A Transfer Entropy and Kramers–Moyal Approach to Financial Networks}

\author[1]{Pouriya Khalilian}[]
\ead{pouriya@email.kntu.ac.ir}
\affiliation[1]{organization={Department of Physics, K. N. Toosi University of Technology, P. O. Box 15875-4416},
city={Tehran}, 
country={Iran}}
            
\author[2]{Amirhossein N. Golestani}[]
\ead{A.naseri@oist.jp}
\affiliation[2]{organization={Okinawa Institute of Science and Technology Graduate University (OIST)},
city={Onna},
country={Japan}} 
	
\author[3]{Mohammad Eslamifar}[]
\ead{mohammad.eslamifar@studenti.unipd.it}
\affiliation[3]{organization={Department of Physics and Astronomy, University of Padua},
city={Padua}, 
country={Italy}} 
	
\author[4]{Mostafa T. Firouzjaee}[]
\ead{taghizade.mostafa@ausmt.ac.ir}
\affiliation[4]{organization={Faculty of Engineering Modern Technologies, Amol University of Special Modern Technologies},
city={Amol}, 
country={Iran}}

\author[1, 5]{Javad T. Firouzjaee}[]
\cormark[1]
\ead{firouzjaee@kntu.ac.ir}
\affiliation[5]{organization={School of Physics, Institute for Research in Fundamental Sciences (IPM), P. O. Box 19395-5531},
city={Tehran}, 
country={Iran}}

\cortext[1]{Corresponding author}

\begin{abstract}
Financial markets are dynamic, interconnected systems where local shocks can trigger widespread instability, challenging portfolio managers and policymakers. Traditional correlation analysis often miss the directionality and temporal dynamics of information flow. To address this, we present a unified framework integrating Transfer Entropy (TE) and the N-dimensional Kramers–Moyal (KM) expansion to map static and time-resolved coupling among four major indices: Nasdaq Composite ($^\wedge IXIC$), WTI crude oil ($WTI$), gold ($GC=F$), and the US Dollar Index ($DX-Y.NYB$). TE captures directional information flow. KM models non-linear stochastic dynamics, revealing interactions often overlooked by linear methods. Using daily data from August 11, 2014, to September 8, 2024, we compute returns, confirm non-stationary using a conduct sliding-window TE and KM analyses. We find that during the COVID-19 pandemic (March–June 2020) and the Russia–Ukraine crisis (Feb–Apr 2022), average TE increases by 35\% and 28\%, respectively, indicating heightened directional flow. Drift coefficients highlight gold–dollar interactions as a persistent safe-haven channel, while oil–equity linkages show regime shifts, weakening under stress and rebounding quickly. Our results expose the shortcomings of linear measures and underscore the value of combining information-theoretic and stochastic drift methods. This approach offers actionable insights for adaptive hedging and informs macro-prudential policy by revealing the evolving architecture of systemic risk.
\end{abstract}

\begin{highlights}
\item Uncovering Hidden Financial Linkages with Transfer Entropy \& Kramers-Moyal Expansion
\item Crises Intensify Information Flow: COVID-19 and Ukraine War Drive Surges
\item Beyond Linear Models: A Tool for Adaptive Hedging and Systemic Risk Policy
\end{highlights}


\maketitle

\section{Introduction}

The intricate connections and dynamic pairwise interactions among economic factors have garnered interest of among scholars, economists, and policymakers. consciousness the correlations between key economic indices, such as oil and gold prices, the US dollar index, and stock prices; provides critical insights into global financial dynamics \cite{intro:0}. Furthermore, quantifying these pairwise-interactions indicate, how different markets influence one another, shaping the broader financial perspective\cite{intro:5}. Pairwise interaction analysis allows the identification of risk exposures and investment opportunities, facilitating informed decision-making \cite{intro:1}.It also supports the strengthen of hedging, diversification, and risk management strategies, thereby enhancing portfolio stability \cite{intro:2}. Furthermore, such analysis offers predictive capabilities, allowing anticipation of market movements based on observed interactions \cite{intro:4}. This is particularly valuable in today’s complex financial systems, where changes in one market can rapidly propagate to others \cite{intro:3}.

Understanding pairwise interactions among crude oil, gold, and the US dollar is essential to gain insights into global economic dynamics and making informed investment decisions. The prices of these commodities and the US dollar index are interlinked in complex ways, influencing one another through various economic and geopolitical factors\cite{intro:5}. For instance, crude oil prices can impact inflation, which in turn affects gold prices as an inflation hedge\cite{intro:3}. Similarly, fluctuations in the US dollar can influence crude oil prices, given that oil is typically priced in dollars, thereby affecting its demand and supply dynamics globally\cite{intro:4}.

Kukreti and et al. conducted an investigation into the correlation-based financial network. The structure of empirical correlation matrices, derived from financial market data, they found that the structure of evolves as individual stock prices fluctuate over time. This evolution is particularly notable during critical events like market crashes and bubbles, exhibiting fascinating patterns\cite{Kukreti}. 

Moreover A.H. Shirazi and et al. introduced a method for mapping random processes onto complex networks. They construct networks for time series data, including the German stock market index (DAX) and white noise networks. They demonstrated that these time series could be reconstructed with high precision through a simple random walk on the corresponding networks\cite{jafari}. 

Arfaoui and et al. employed a system of simultaneous equations to identify the direct and indirect connections among these economic assets over a certain period. The study found negative relations between oil prices and stocks, oil prices and the US dollar rate, and the US dollar rate and gold prices. Additionally, they noted a close relation between the price of oil and the price of gold\cite{arfaoui2017oil}.

Juárez and et al\cite{juarez2011applying}, concluded that the relation between gold reserves, the price of energy products, and the spot price of gold is positive. In addition, the study determined that financial indicators negatively affect the gold spot price. This finding aligns with those in Refs.\cite{arfaoui2017oil, shen2014us}, indicating that during times of crisis and recession, shareholders prefer gold as a safe haven. Conversely, during economic growth, they leave the gold market and re-enter the capital market in hopes of higher profits. Therefore, this negative relation is reasonable and applies under normal conditions. Furthermore, the author claims a negative correlation between global macroeconomic data and gold spot prices. The price of gold and the inflation rate are interrelated, as both variables significantly impact people's daily lives.

In addition, the relation between oil prices and exchange rates, and why changes in oil prices should affect exchange rates, have been explored by Globe and Krugman\cite{golub1983oil, krugman1983oil}. Globe argued that because oil is priced in dollars, an increase in the price of oil will boost the demand for US dollar. However, Krugman focused on the relation between portfolio investment preferences of oil exporters and exchange rate movements. He asserted that a change in the price of oil will lead to more activities in the portfolios of oil exporters. Krugman believed that exchange rates are primarily determined by the movement of the current account, and consequently, if a raise in oil price leads to a weakening of a country's current account, the exchange rate will decrease. Many other studies indicated that the impact of US dollar index on the international price of oil is significant in the long run. Moreover, monetary policy and interest rate changes may also contribute to the negative correlation among exchange rates and prices.

Numerous pieces of evidence demonstrated that the modification use of oil as a financial asset over the past decades had intensified its relation with other assets. However, characteristics such as the chaotic and non-linear nature of crude oil time series have made accurate price estimation challenging. To address these issues, many economists and scientists have developed new models for predicting crude oil prices. For instance, Karasu and Altan proposed a model that comprised Long Short-Term Memory (LSTM) networks, technical indicators such as trend, volatility, momentum and the Chaotic Henry Gas Solubility Optimization (CHGSO) technique. In their model, they also utilized features based on technical indicators of trend, momentum, and fluctuations. They separately derived these indices for West Texas Intermediate (WTI) and Brent Crude Oil \cite{karasu2022crude, Firouzjaee}.

In the foreign exchange market, exchange rate analysis are typically based on the overall health of a country's economy. Generally, exchange rate for a currency of a country with growing economy reflects its higher value. Establishing a straightforward relation across currencies using a common commodity is not feasible, as the interconnection between price of gold, exchange rates, and oil price is complex. Various factors can influence the prices of gold and crude oil, including governmental policies, budget size, inflation, national economic and political conditions, among others\cite{sujit2011study}.

In this research, we initially analyze the descriptive statistics of the pairwise relations between main economic indices and the Nasdaq. Subsequently, we present methods to examine these pairwise interactions, such as the Kramers-Moyal expansion and the Transfer entropy. Lastly, we interpret the findings to gain a deeper understanding of the changing dynamic interactions during the Ukraine crisis and the COVID-19 pandemic.

\section{Related Works}

The analysis of financial markets increasingly exploits information-theoretic and stochastic modeling to capture complex, time-varying interactions among assets. Transfer Entropy (TE), introduced by Schreiber \cite{ds:12}, quantifies directed information flow between time series and has been widely adopted to study causality and contagion in markets. Early the Econophysics applications by Kwon and Yang \cite{rw:1} and Marschinski and Kantz \cite{rw:2} demonstrated that TE uncovers asymmetric dependencies among global equities and foreign-exchange indices. Subsequent refinements addressed methodological nuances: Kyrtsou et al. \cite{rw:3} developed an asymmetric causality test for S\&P 500–VIX volatility–volume linkages; Dimpfl and Peter \cite{rw:4} proposed group TE to measure joint volatility spillovers across cryptocurrencies; and Yang and Shang \cite{rw:5}, as well as He and Shang \cite{rw:6}, systematically compared TE estimators on financial data. More recent work employs sliding-window TE to build time-resolved networks: Gao et al. \cite{rw:7} analyzed Chinese A-shares during the 2008 crisis to identify sectoral contagion hubs; Peng et al. \cite{rw:8} mapped temporal shifts in Chinese sectoral capital flows; Zhao et al. \cite{rw:9} constructed rolling TE networks of commodity futures to provide early systemic-risk warnings; and Janczewski et al. \cite{rw:110} used conditional TE on high-frequency FX-dealer data around an ECB announcement, revealing transient network motifs. Extensions include Liu et al. \cite{rw:20}, who applied TE to link economic-policy uncertainty, investor sentiment, and stock returns, and Wang et al. \cite{rw:22}, who examined entropy co-movement across global markets.

Complementing TE, the Kramers–Moyal (KM) expansion underlies stochastic drift-diffusion modeling of financial time series, yielding drift $D^{(1)}$ (x,t) and diffusion $D^{(2)} (x,t)$ functions via empirical probability moments. Ivanova et al. \cite{rw:11} applied a Fokker–Planck framework to equity and FX returns, while Friedrich et al. \cite{rw:12} extended KM methods to account for jump-diffusion processes in markets exhibiting fat tails. Despite providing mechanistic insights into single-series dynamics, KM approaches traditionally assume Markovian behavior and do not capture cross-asset information flows or network structures.

Complex-network and entropy-based studies have further enriched systemic-risk analysis. Correlation-based minimum spanning trees reveal core–periphery architectures in equity markets \cite{rw:13}, whereas information-theoretic networks—using mutual information or TE—detect nonlinear, directed relationships. Bekiros et al. \cite{rw:14} contrasted correlation and TE networks across U.S. equities and commodities during 2008, observing distinct community reorganization; Cerchiello and Aste \cite{rw:15} built daily TE and mutual information graphs for U.S. banks, linking network metrics to stress indicators; and Goh et al. \cite{rw:16} inferred small-world equity-FX networks via normalized mutual information. Machine-learning approaches have also emerged: Rakib et al. \cite{rw:17} used feature-ranking to track evolving stock networks, and Karpman et al. \cite{rw:18} applied random forests to intraday returns, detecting pre-crisis connectivity surges.

Nevertheless, the literature remains bifurcated: TE-based models excel at mapping where and how strongly information propagates but lack explicit dynamic laws, while KM-based methods describe how individual series evolve but ignore network effects. Furthermore, most studies focus on a single asset class or geography, and even multi-asset analyses (e.g., Goh et al. \cite{rw:18}) do not integrate stochastic-drift information. To our knowledge, no prior work unifies TE-driven network analysis with N-dimensional KM expansion in a multi-asset framework. This paper fills that gap, offering a coherent methodology to track directional information flow and stochastic dynamics across equity, commodity, and currency markets under both normal and crisis conditions.

\section{Data and descriptive statistics}

\begin{figure}[h!]
	\centering
	\includegraphics [width=0.4 \columnwidth] {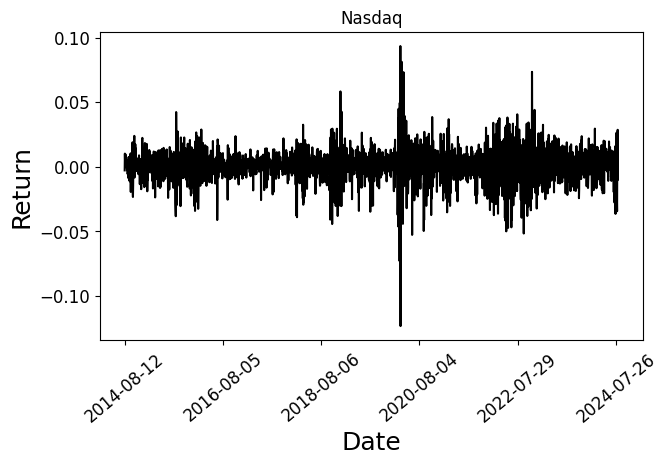}
	\includegraphics [width=0.4 \columnwidth] {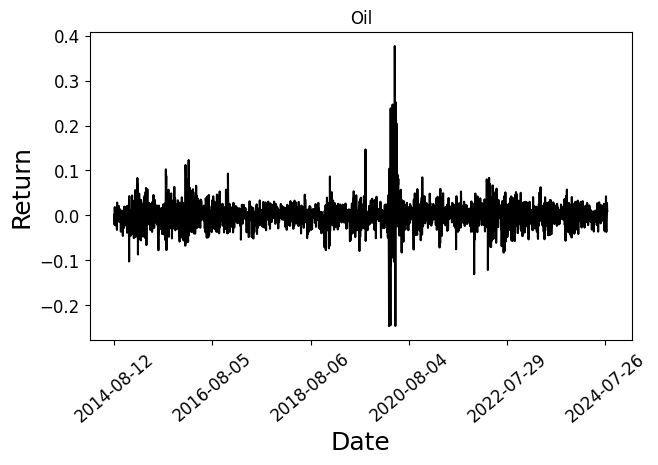}
	\includegraphics [width=0.4 \columnwidth] {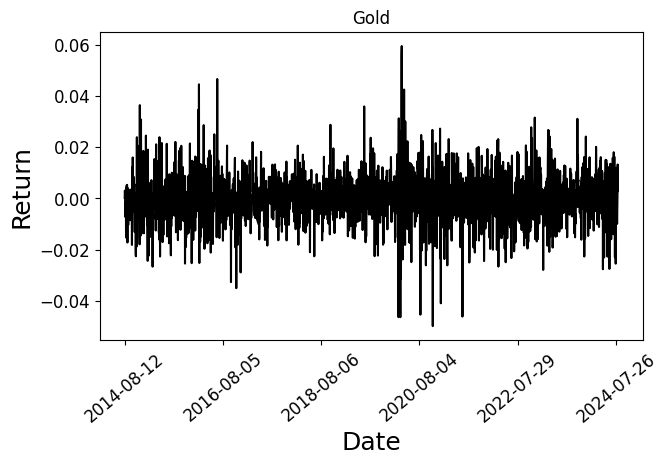}
	\includegraphics [width=0.4 \columnwidth] {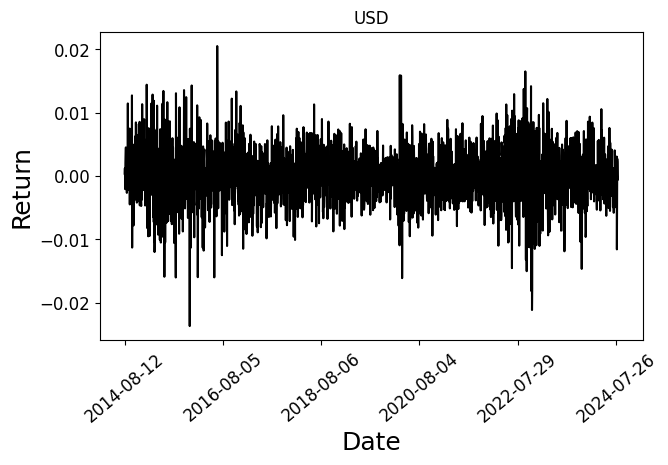}
	\caption{\scriptsize{Daily stock returns for financial indices.}}
	\label{fig0:0}
\end{figure}

In this study, we have applied daily stocks price indices data for four main financial markets. The data set includes Nasdaq index ($^\wedge IXIC$), crude oil index ($CL=F$), global gold index ($GC=F$), and US dollar index ($DX-Y.NYB$). All data is obtained from yahoo finance api\cite{yfapi}. The summary of statistics of stock-index returns (the returns are expressed as percentages) in the four indices are presented in Table \eqref{table0:0}. To visualize the returns for each market, we depict the series in Fig.\eqref{fig0:0}. 

\begin{table} [h!]
	\centering
	\begin{tabular}{ | c | c | c | c | c | }
		\hline
		& \textbf{Mean} & \textbf{Standard deviation} & \textbf{ Skewness} & \textbf{Kurtosis} \\ [0.5 ex]
		\hline
		\textbf{Nasdaq}    & 0.00062 & 0.0133 & -0.4417 & 7.5326 \\\hline
		\textbf{Crude-oil} & 0.00057 & 0.0294 & 1.1527  & 25.138  \\\hline
		\textbf{Gold}      & 0.00029 & 0.0092 & 0.0086  & 3.7718  \\\hline
		\textbf{US-dollar} & 0.0001  & 0.0043 & -0.0757 & 1.6591 \\[1ex]
		\hline
	\end{tabular}
	\caption{Descriptive statistics on stock returns from 11/08/2014 to 08/09/2024. Observations for all series in the whole sample period are 2513.}
	\label {table0:0}
\end{table}

A notable statistic of the stock return series shown in Table\eqref{table0:0} contain the high value of kurtosis for crude-oil. This suggests that, for these markets, big shocks of either sign are more likely to be present and please omit that the stock-return series may not be normally distributed. One of the things that always influence on the price of oil is the amount of production capacity and the amount of oil entering the market, in this regard Ref. \cite{dees2008assessing} expressed that the existence of a non-linear relation between the price of oil and the amount of oil that has been supplied to the market is possible to affect the determination of oil prices. The outbreak of the Covid-19 pandemic has caused unusual volatility for stocks in early 2020. The main reason for this unusual fluctuation can be related to investor sentiments\cite{news}, on the other hand, investor sentiment in the energy market significantly increased during tail events, suggesting that anxious investors flocked to put options and paid a higher premium to shield themselves from unprecedented risks in the energy market\cite{eng:1}. By examining the collective behavior of people based on social networks, it could be seen that the Nasdaq market was affected by investors during the Covid-19 pandemic and had abnormal fluctuations\cite{eng:2}. According to the literature, gold is considered as a safe asset for investors, and for this reason, unlike oil and Nasdaq, it is not unusually fluctuated\cite{eng:3}.
The results of Ref. \cite{dees2008assessing} show the effect of utilization rate for the refinery on price of crude oil, which demonstrates that a decrease in the refining rate leads to an change in the price of oil.
However, its results indicate that there is limited evidence so that increasing the capacity of refineries can reduce oil prices. According to their research, only the crude oil can effectively reduce the price.
Moreover Ref. \cite{nyangarika2019oil} concluded, that by happening the Iraq war in 2004, that military operations in oil-producing countries have a negligible effect on oil prices. However, this conclusion can be contested in light of the recent Ukraine-Russia war, which caused significant volatility in oil prices. Although this fluctuation may be temporary, it underscores the importance of the military operations on oil prices \cite{firouzjaee2022machine}. In addition, gold remain an influencing index in the market.

Ref. \cite{shen2014us} stated that gold has long been appreciated and has been considered as key a financial tool. It can be viewed as a resource so that is independent of all nations and industries. According to the author, many who possess gold think suppose that investing in the gold market can provide protection against inflation. Moreover, Ref. \cite{arfaoui2017oil} further underlined that there is ample proof between gold price rises and stock prices fall during a period of economic turbulence. All eyes are on gold in this situation as a haven of safety. On the other hand, gold also plays the role of a valuable financial asset or economic security against economical volatility. With the occurrence of Covid-19 pandemic crisis, the price of gold faced fluctuations of 26\% in 2020. Therefore, it can be inferred that fluctuations in the prices of gold and crude oil are connected to macroeconomic structures and business cycles \cite{guan2021volatility}.

A recent data set from 2000 to 2020 includes two crises and critical economic situations (the global economical crisis in 2008 and the ongoing Covid-19 pandemic in 2020), showing that in July 2008, oil prices reached their highest level of 145.18 US dollars per barrel. However, with the outbreak of pandemic, the price of oil reached an all-time low of \$37.63 per barrel. On the other hand, even though the price of crude oil was falling during the pandemic, gold price rose dramatically, reached to a peak of \$2,060 per ounce in August 2020. The following months, gold price  fell sharply below \$1,800 per ounce in November 2020. These fluctuations for Gold were more than those expected. Volatility and movement of prices in the last two decades' crisis have prompted economists to examine effects on oil and gold prices on the economic growth of resource-producing countries \cite{guan2021volatility}.

Ref. \cite{oloko2021fractional} indicated that pairwise interactions of the gold price shock remains for a long time on the inflation rate of developing countries and for a short period on the inflation rate of developed countries. The conclusion of Ref. \cite{tkacz2007gold} about gold prices and inflation rates indicates that gold contains significant information regarding future inflectional of various countries, especially for those which have adopted official inflation targets.	Gold is a high-liquidity asset that allows holders to exchange at any time. Many individuals see investing in the gold market as a hedge against inflation and currency depreciation. From the economic and financial perspective, gold price movements are highly regarded.

Historically, despite the losses that occurred during periods of inflation, social unrest, and war, the price of gold changed little. As stock prices fall, financial advisors suggest the investors to maintain investments in gold during this period. On the contrary, some investors believe that during the boom period, the price of gold does not have a risk-averse portfolio value and it can be considered like other commodities whose price swings are based on supply and demand. As a result, in this case, gold investments often fall in value as stock prices rise. In the context of stock market's inverse relation with oil price volatility, oil investments also function as a risk deterrent \cite{sujit2011study}.

The Nasdaq stock market is an American stock exchange that ranks second in the list of stock exchanges in terms of traded stock market value. 
The dynamic pairwise interactions between the Nasdaq and main economic indices such as oil, gold, and the US dollar have significant implications for investors and economist. Meanwhile, gold, often seen as a safe haven asset, tends to move inversely with stock indices such as the Nasdaq during periods of market uncertainty, reflecting future risks for investor sentiment\cite{nas:0}. Furthermore, the value of the US dollar can affect multinational companies listed on the Nasdaq, influencing on their international earnings and competitiveness\cite{nas:1}.

\section{Quantitative Methods}

A fundamental challenge in financial networks is characterizing the complex dynamics of the nodes by measuring the interactions in the stock market. Various methods have been presented to investigate complex relation in time series, although they may not offer comprehensive integration. For instance, Olive M. Cliff et al.\cite{int:0} have classified these methods into six categories: 1) Basic statistical method \cite{int:1}, 2) Distance similarity\cite{int:2, int:3, int:4, int:5}, 3) Causal indices\cite{int:6}, 4) Information theory\cite{int:7, int:8, ds:12, int:10}, 5) Spectral methods\cite{int:11, int:12}, and 6) Miscellaneous method such as co-integration\cite{int:13} and model fit.

Generally, various methods have been proposed to investigate network and pairwise interactions in complex systems\cite{ds:3, ds:4}, each with its own advantages and disadvantages\cite{ds:1, ds:2}. Financial markets can be described as complex systems in which different stocks, including key economic indices, interact and exhibit collective behaviors\cite{ds:5}. The correlation coefficient and mutual entropy are methods used to examine pairwise interactions in the stock market as instantaneous and symmetric interaction\cite{ds:6, ds:7}. In addition, transfer entropy\cite{ds:12} and the Kramers-Moyal expansion are used to analyze the dynamic evolution in time series data, allowing the derivation of dynamic interactions\cite{ds:10, ds:8}.
\subsection*{Mutual entropy and Transfer entropy}

In pairwise systems, crucial information about interactions can be taken via measuring the extent of components' participation to generate information and exchange the degree of information. Although mutual entropy can determine information between two pairwise systems, it lacks dynamic information and the direction of information flow. Transfer entropy incorporates some of the desired features of mutual entropy and additionally includes the dynamics of information transfer. Mutual entropy is defined as follows:\cite{ds:11}: 
\begin{gather}\label{eq1:1}
	I_{ij} = \sum_{i,j} P(i,j) \log_2{\frac{P(i,j)}{P(i) P(j)}}
\end{gather}
where it is a method used to assess the degree of independence between two processes. Here $I_{ij}$ is symmetric in the exchange of information between i and j, and lacks direction. By introducing a time delay in each variable, the dynamic direction may be  established, which allows us to calculate the transfer entropy\cite{ds:12}.
\begin{gather}\label{eq2:1}
	T_{ji} = \sum_{i,j} P(i_{t+dt},i_t,j_t) \log_2{\frac{P(i_{t+dt}|i_t,j_t)}{(i_{t+dt}|i_t)}}
\end{gather}

Eventually, the transfer entropy detects the direct exchange of information between two pairwise systems. For N pairwise systems, the transfer entropy matrix is as follows:	

\begin{gather}
	T
	=
	\begin{bmatrix}
		T_{11} & \ldots & T_{1N} \\
		\vdots & \ddots & \vdots \\
		T_{N1} & \ldots & T_{NN} \\	
	\end{bmatrix}
\end{gather}

\subsection*{Pairwise interactions using $N$-dimensional Kramers-Moyal expansion}
Pairwise interaction for a complex system can be expressed as a linear set of differential equations as\cite{ds:9}:
\begin{gather}\label{eq:1}
	\frac{d}{dt}x(t) = A x(t)
\end{gather}
where, 
\begin{equation*}
	x(t) \in \mathbb{R}\ , \: and \: A \: is \: N {\times} N matrix
\end{equation*}
each of elements in interaction matrix, $A_{ij}$, measures the effect of a slight change in the value of variable $x_j$ on $x_i$. The derivation of stochastic differential equations to determine dynamic equation from time series data, involves analyzing multiple time series in a pairwise manner. One may yield linear effects of such interactions from equation (\ref{eq:1}), referred to as Drift, which is the first term of the Kramers-Moyal expansion:
\begin{gather}\label{eq:2}
	D^{(1)}(x,t)=\lim_{dt\to\infty} \frac{1}{dt}  <x_i (t+dt)-x_i>|_{(x(t)=x=x_1,x_2,\ldots,x_n)}\     	
\end{gather}
\begin{equation*}\label{eq:3}
	=\lim_{dt\to\infty} \frac{1}{dt}  \int\ dx^\prime\ p(x^\prime,t+dt|x,t)\ (x_i^\prime-x_i)  	
\end{equation*}
where $dt$ is time delay and $<\ldots>$  is conditional averaging.

Drift is determined from the conditional probability of $p(x^\prime,t+dt|x,t)$ at small time delay $dt$. We can set equation(\ref{eq:1}) equal to $D^{(1)}(x,t)$ as:
\begin{gather}\label{eq:5}
	\frac{d}{dt}x(t) = A x(t) = D^{(1)}(x,t)
\end{gather}
As a result, drift can be derived from equation(\ref{eq:2}) and written as:
\begin{gather}\label{eq:6}
	<x_i (t+dt)-x_i(t)>|_{(x(t): x : x_1,x_2,\ldots,x_n)}\  =  <y_i (t,dt)>|_{(x(t)=x=x_1,x_2,\ldots,x_n)}\ 	
\end{gather}
\begin{equation*}\label{eq:7}
	=\psi_{i,1}x_1+\psi_{i,2}x_2+\ldots+\psi_{i,N}x_N
\end{equation*}
where,
\begin{equation*}\label{eq:8}
	\psi_{l,j}=\psi_{l,j}(dt) = dt A_{ij} ;\text{    } i,j = 1,2,3, \ldots,N
\end{equation*}

the conditional averaging is defined as follows, and could apply Bayes theorem in the following:
\begin{gather}\label{eq:9}
	<y_i (t,dt)>|_{(x(t)=x=x_1,x_2,\ldots,x_n)} = \int y_i p(y_i |x_1 x_2 \ldots x_n)dy_i =\int y_i \frac{p(y_i,x_1,x_2,\ldots,x_n)}{p(x_1,x_2,\ldots,x_n)}dy_i	
\end{gather}
\begin{equation*}\label{eq:10}
	=\psi_{i,1}x_1+\psi_{i,2}x_2+\ldots+\psi_{i,N}x_N	
\end{equation*}
then, may yield:
\begin{gather}\label{eq:11}
	<y_i(t,dt)x_j>=\psi_{i,1}<x_1 x_j>+\ldots+\psi_{i,N}<x_N x_j>	
\end{gather}
One may note that the averaging in the whole expression and in equation(\ref{eq:11}) does not depend on any condition of x(t), so we may derive the coefficients of $\psi_{i,j}$ as follows:

\begin{gather}
	\begin{bmatrix}
		<y_i(t,dt)x_1>  \\ 
		<y_i(t,dt)x_2> \\
		\vdots \\ 
		<y_i(t,dt)x_N> 
	\end{bmatrix}
	=
	\begin{bmatrix}
		<x^2_1> & \ldots & <x_1 x_N> \\
		\vdots & \ddots & \vdots \\
		<x_N x_1> & \ldots & <x^2_N> \\	
	\end{bmatrix}
	\begin{bmatrix}
		\psi_{i,1}  \\ 
		\psi_{i,2} \\
		\vdots \\ 
		\psi_{i,N}
	\end{bmatrix}
\end{gather}

\section{Results and discussion}
\begin{figure}[h!]
	\centering
	\includegraphics [width=0.4 \columnwidth] {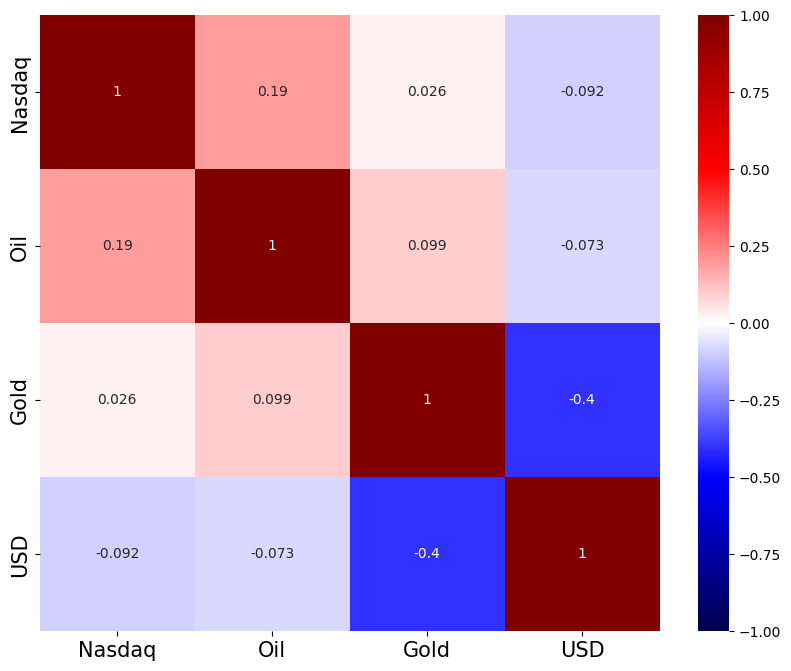}
	\includegraphics [width=0.4 \columnwidth] {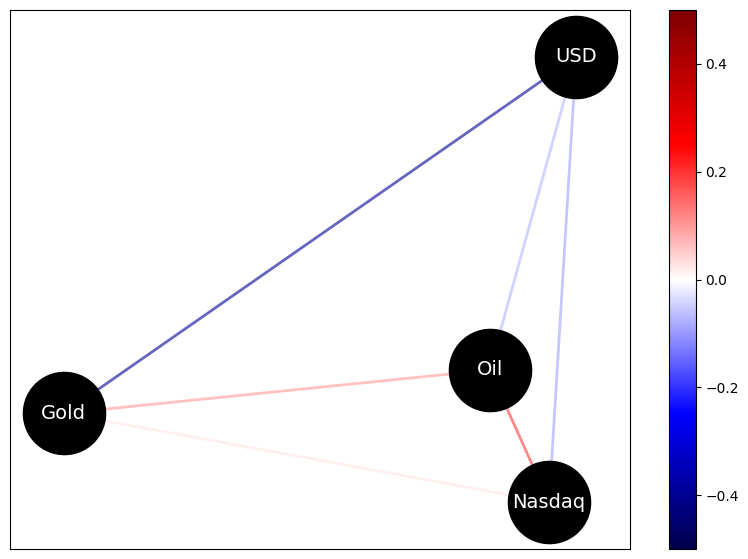}
	\caption{\scriptsize{Correlation coefficient heatmap and symmetric network for crude oil, US dollar, gold and Nasdaq.}}
	\label{fig:1}
\end{figure}

\begin{figure}[h!]
	\centering
	\includegraphics [width=0.9 \columnwidth] {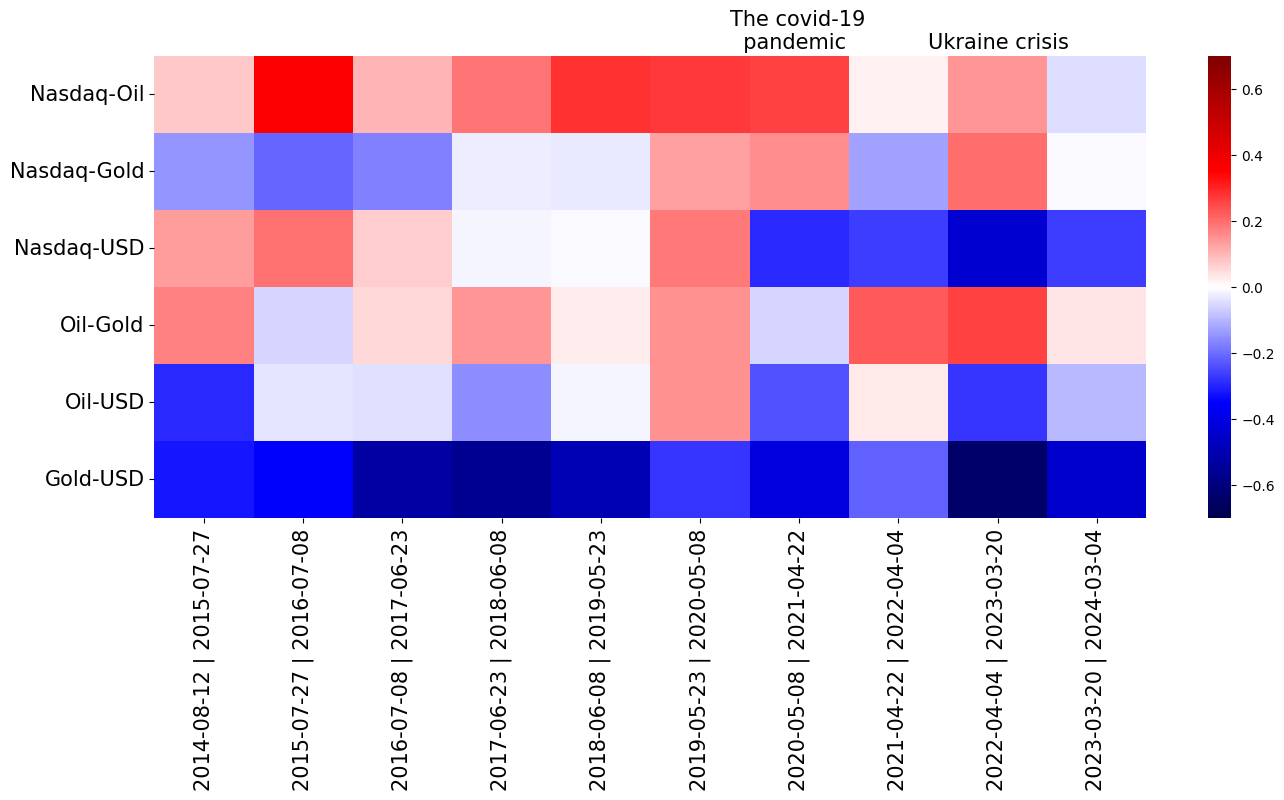}
	\caption{\scriptsize{Correlation coefficient evolution heatmap.}}
	\label{fig:ec}
\end{figure}

First, to investigate the impact of crude oil, gold, US dollar index, and Nasdaq index pairwise interaction network, we have used daily time step data from 2014-08-11 to 2024-09-08. According to Fig. \eqref{fig:1} and Table \eqref{table:1}, one may illustrate the symmetric pairwise interaction network so that is obtained utilizing correlation coefficient, the results are presented in figures \eqref{fig:1} \eqref{fig:ec}:

\begin{table} [h!]
	\centering
	\begin{tabular}{ | c | c | c | c | c | }
		\hline
		& \textbf{Nasdaq} & \textbf{Crude-oil} & \textbf{Gold} & \textbf{US-dollar} \\ [0.5 ex]
		\hline
		\textbf{Nasdaq}    & 1 & 0.19 & 0.026 & -0.091 \\\hline
		\textbf{Crude-oil} & 0.19 & 1 & 0.099 & -0.07  \\\hline
		\textbf{Gold}      & 0.026 & 0.099 & 1 & -0.4  \\\hline
		\textbf{US-dollar} & -0.091 & -0.07 & -0.4 &1 \\[1ex]
		\hline
	\end{tabular}
	\caption{Effective correlation coefficient results.}
	\label {table:1}
\end{table}

The results indicate that Nasdaq index has only a weak negative linear relation with crude oil, suggesting that fluctuations in the Nasdaq index are minimally influenced by changes in crude oil prices. In addition, gold exhibits a weak linear relation with US dollar index, indicating variations in gold prices are only slightly impacted by movements in US dollar index. According to the correlation coefficient analysis, the rest of pairwise relations demonstrate independence, meaning that there are no significant linear relations among other indices. This independence suggests that the interactions among these pairs do not exhibit a straightforward linear relation, indicating more complex underlying dynamics that might not be captured by linear correlation alone. For a better understanding of outliers and highlights, a scatter plot has shown in Fig. \eqref{fig:2}:

\begin{figure}[h!]
	\centering
	\includegraphics [width=0.3 \columnwidth] {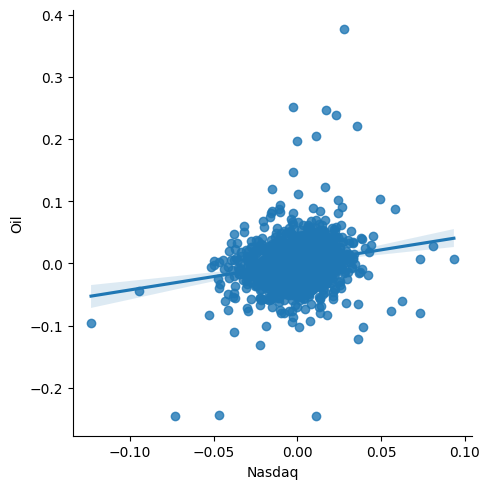}
	\includegraphics [width=0.3 \columnwidth] {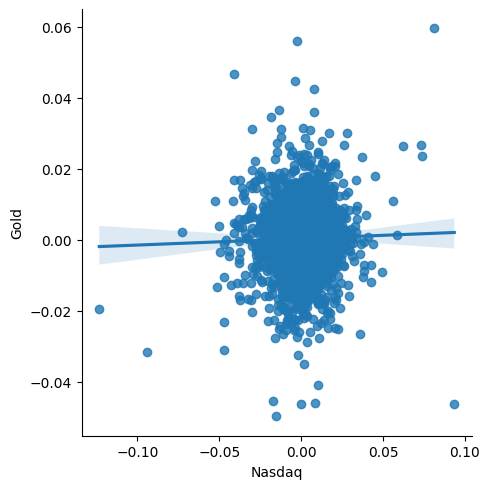}
	\includegraphics [width=0.3 \columnwidth] {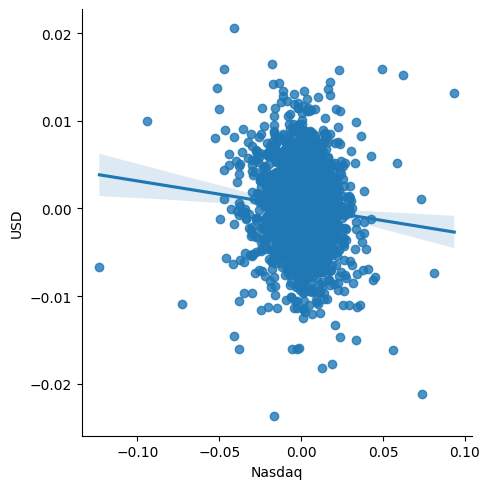}
	\includegraphics [width=0.3 \columnwidth] {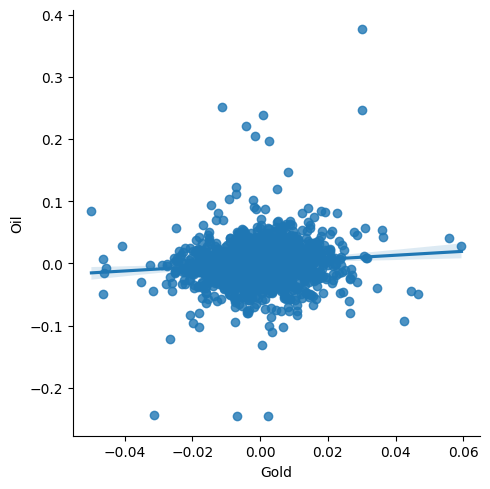}
	\includegraphics [width=0.3 \columnwidth] {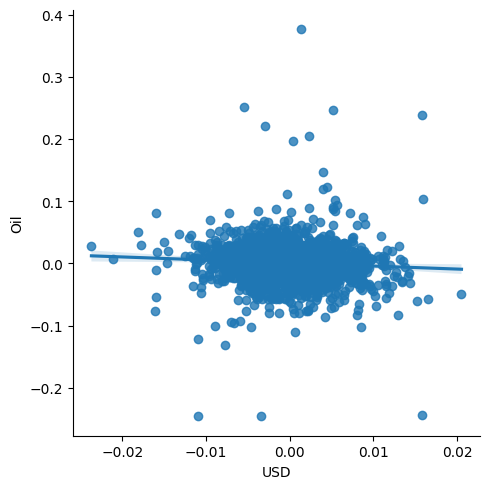}
	\includegraphics [width=0.3 \columnwidth] {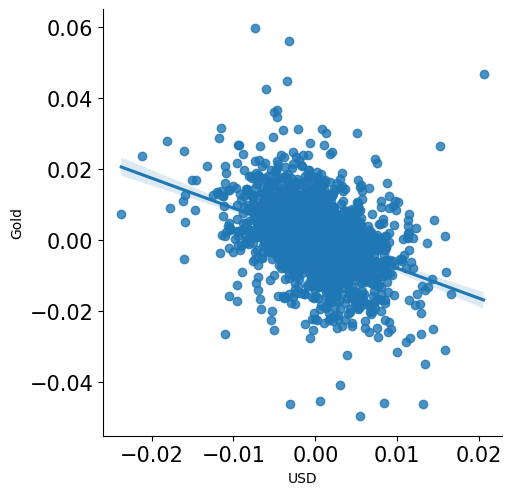}
	\caption{\scriptsize{This scatter plot illustrates the correlation between the Nasdaq index and other financial indices such as crude oil, gold, and the US dollar. Each point represents paired data points from the Nasdaq index and the respective asset, allowing visualization of their relation and correlation strength.}}
	\label{fig:2}
\end{figure}

Also used mutual entropy to investigate non-linear relations based on the information generation among indices. The mutual entropy analysis provides valuable insights into the information exchange among different financial indices. By measuring the information using mutual entropy, one can understand  the degree of dependency and the strength of relations. As shown in Table \eqref{table:mi}, and Fig. \eqref{fig:mi}, the results and network derived from information theory are presented as follows:

\begin{table} [h!]
	\centering
	\begin{tabular}{ | c | c | c | c | c | }
		\hline
		& \textbf{Nasdaq} & \textbf{Crude-oil} & \textbf{Gold} & \textbf{US-dollar} \\ [0.5 ex]
		\hline
		\textbf{Nasdaq}    & 1.8 & 0.068 & 0.062 & 0.076 \\\hline
		\textbf{Crude-oil} & 0.068 & 1.44 & 0.065 & 0.058  \\\hline
		\textbf{Gold}      & 0.062 & 0.066 & 2.1 & 0.18  \\\hline
		\textbf{US-dollar} & 0.076 & 0.058 & 0.18 & 2.3 \\[1ex]
		\hline
	\end{tabular}
	\caption{Mutual information results.}
	\label {table:mi}
\end{table}

The results show that, the information values highlight significant interaction among specific pairs of indices. The information between gold and the US dollar has shown a substantial exchange, which indicates fluctuations in gold have a considerable impact on the value of US dollar and vice versa. The information among other indices represent that, while lower than gold and US dollar pairs, still there is a meaningful interaction, emphasizing the inter-dependency of these financial indices. According to the amount of information taken from stocks' self-interaction, it can e concluded that gold and US dollar have more instability and chaos.

As a result, the mutual entropy results underscore the complex dynamics and interconnection within the financial markets, provide a nuanced understanding that complements the linear correlations observed.

\begin{figure}[h!]
	\centering
	\includegraphics [width=0.4 \columnwidth] {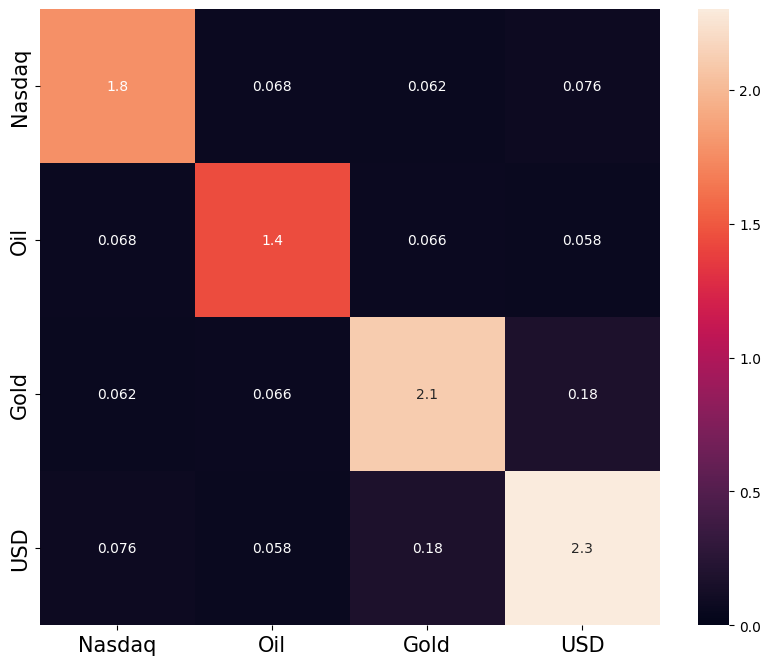}
	\includegraphics [width=0.4 \columnwidth] {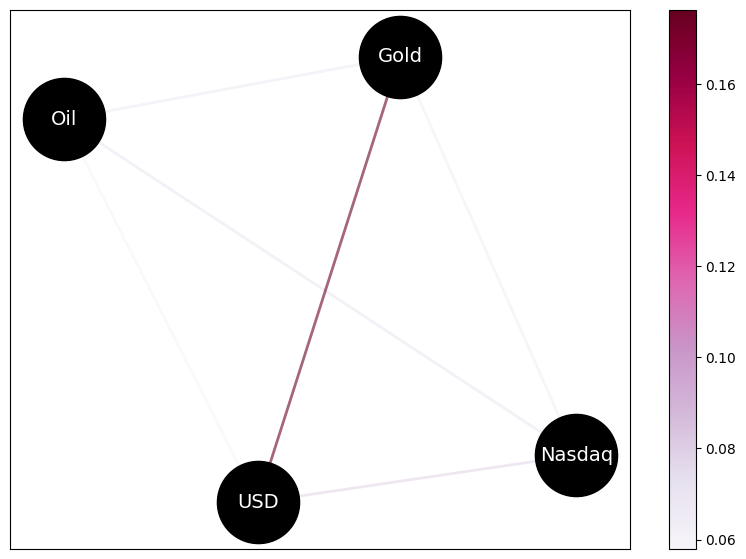}
	\caption{\scriptsize{Mutual information heatmap and symmetric network for crude oil, US dollar, gold and Nasdaq.}}
	\label{fig:mi}
\end{figure}

\begin{figure}[h!]
	\centering
	\includegraphics [width=0.9 \columnwidth] {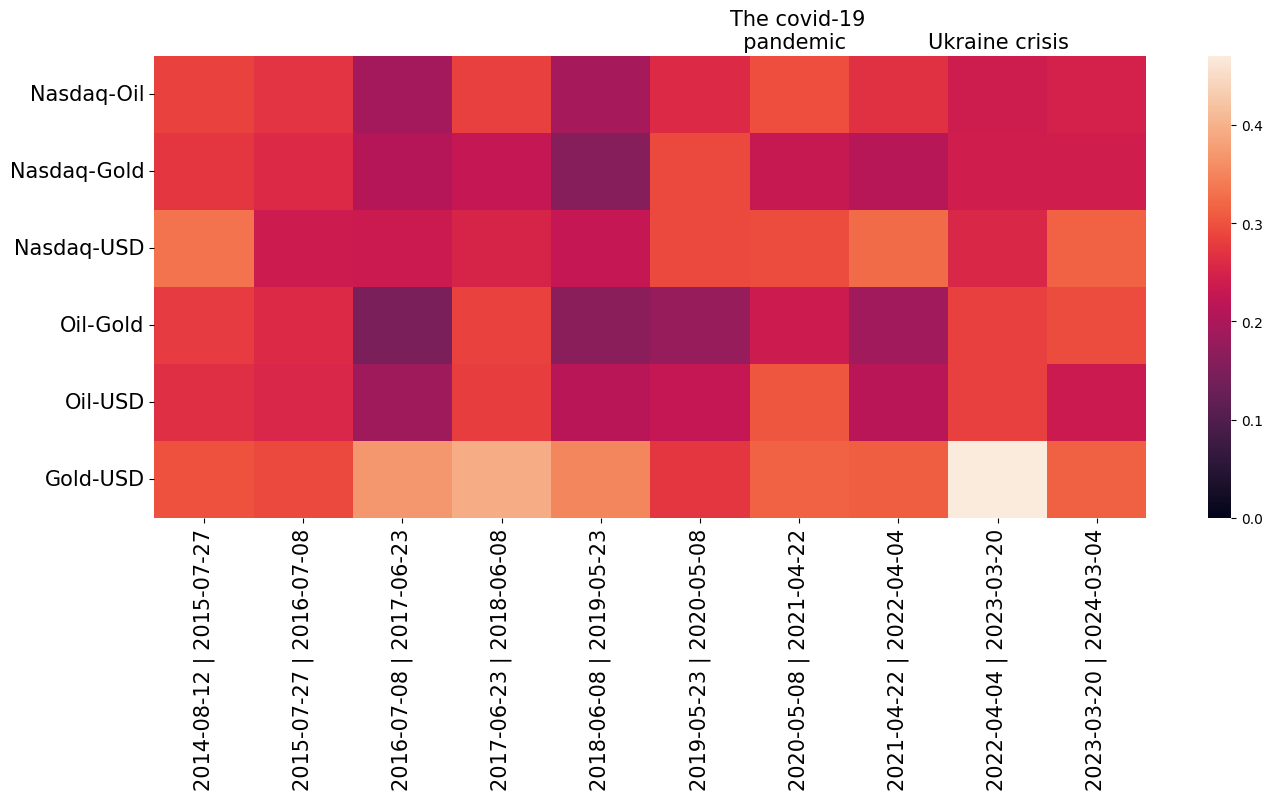}
	\caption{\scriptsize{Mutual information evolution heatmap.}}
	\label{fig:mi}
\end{figure}

\begin{table} [h!]
	\centering
	\begin{tabular}{ | c | c | c | c | c | }
		\hline
		& \textbf{Nasdaq} & \textbf{Crude-oil} & \textbf{Gold} & \textbf{US-dollar} \\ [0.5 ex]
		\hline
		\textbf{Nasdaq}    & 1.7 & 0.021 & 0.0075 & 0.01 \\\hline
		\textbf{Crude-oil} & 0.019 & 1.4 & 0.017 & 0.011  \\\hline
		\textbf{Gold}      & -0.0004 & 0.016 & 2 & 0.11  \\\hline
		\textbf{US-dollar} & 0.0097 & 0.01 & 0.11 & 2.2 \\[1ex]
		\hline
	\end{tabular}
	\caption{Transfer entropy results.}
	\label {table:2}
\end{table}

\begin{figure}[h!]
	\centering
	\includegraphics [width=0.4 \columnwidth] {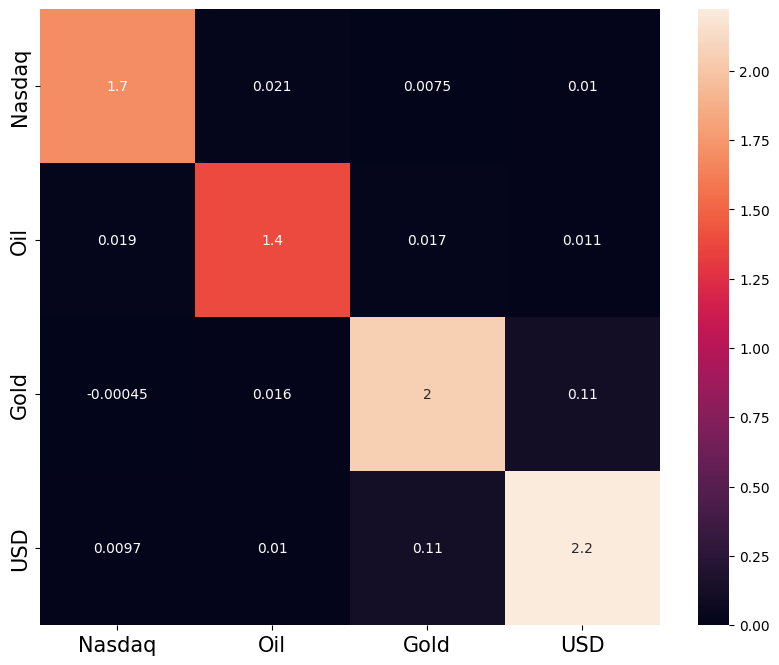}
	\includegraphics [width=0.4 \columnwidth] {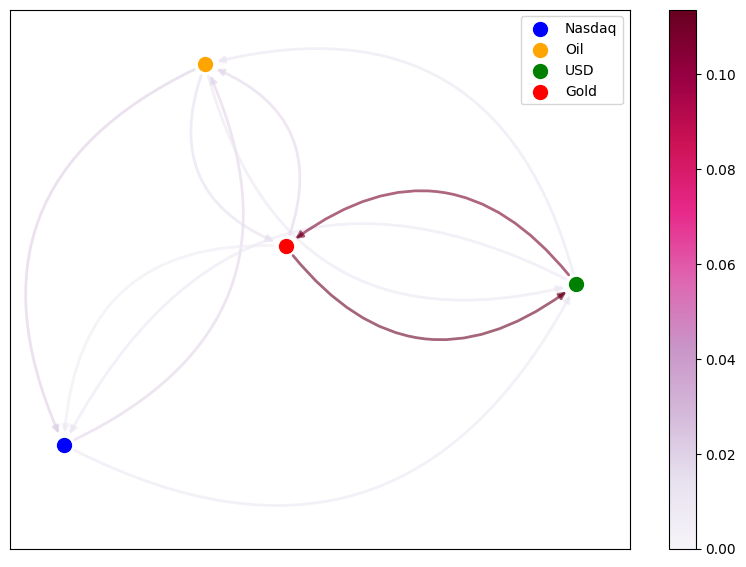}
	\caption{\scriptsize{Transfer information heatmap and network for crude oil, US dollar, gold and Nasdaq.}}
	\label{fig:3}
\end{figure}

Transfer entropy results from Table \eqref{table:2} and Fig. \eqref{fig:3} conclude that US dollar index and gold mutual relation, possess more symmetric information exchange. Finding also has shown one-way flow from Nasdaq index to gold. Other information flows are relatively weaker, compared to above.

Likewise, Fig. \eqref{fig:6} illustrates market's information flow evolution in terms of time. For instance, although Covid-19 pandemic had previously caused to a minor boost in US dollar and gold information flow, it had obviously risen during Ukraine crisis and behaved in a more chaotic way. Meanwhile other pairwise interactions exhibit weaker flow.  Figs. \eqref{fig:ec} and \eqref{fig:mi} also contain similar results about Ukraine crisis and covid-19 pandemic.

\begin{figure}[h!]
	\centering
	\includegraphics [width=0.9 \columnwidth] {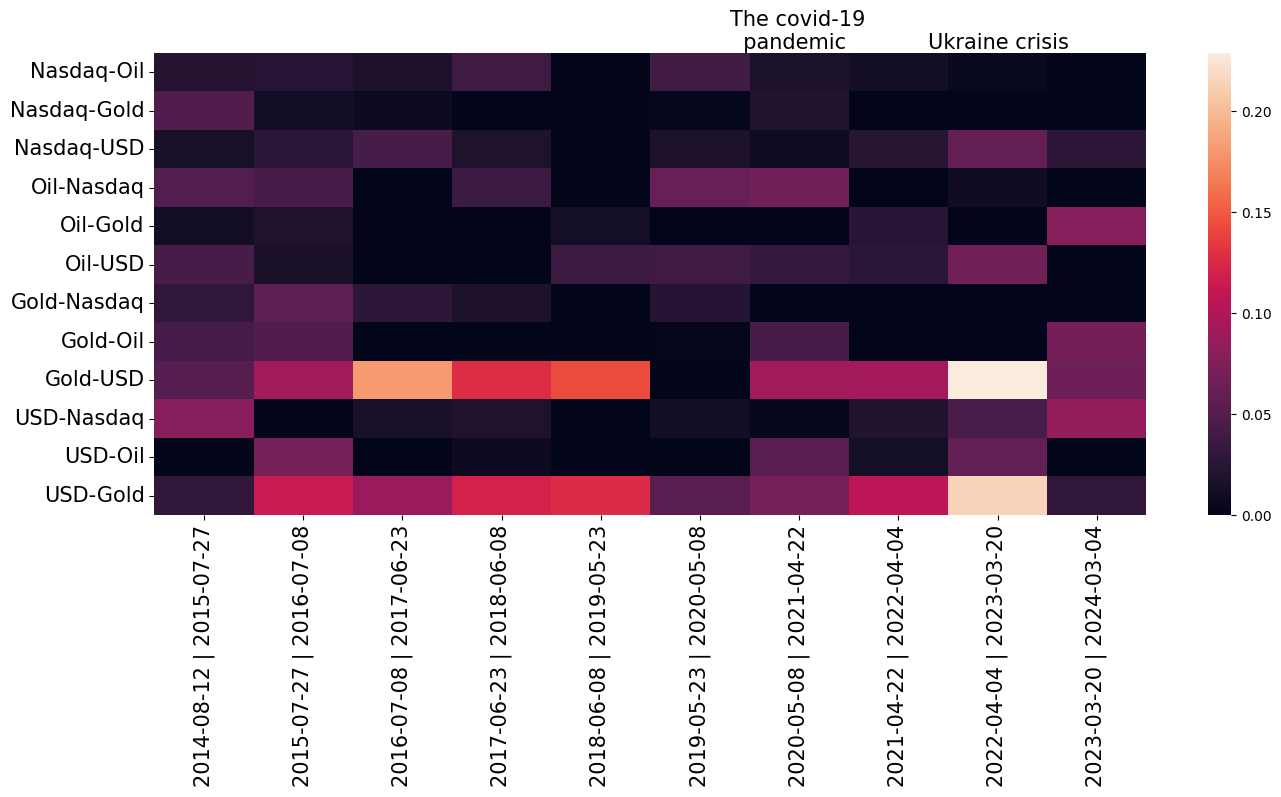}
	\caption{\scriptsize{Transfer information evolution heatmap.}}
	\label{fig:6}
\end{figure}

Obtained interaction coefficients show the relation among price of each stock and others. Figs. \eqref{fig:5} ,\eqref{fig:6} and Table. \eqref{table:3} clearly show the received effects of each stock from other stocks in its growth, aligned or non-aligned. vertical coefficients for each stocks show the influence of each stock on others. Likewise, the horizontal coefficients illustrated in front of each stock indicate the stocks' impact on others. The negative values in diagonals of matrix indicate the non-divergent fluctuation of each stock. Based on the drift coefficient, we may analyze the market's dynamic behavior.

\begin{table} [h!]
	\centering
	\begin{tabular}{ | c | c | c | c | c | }
		\hline
		& \textbf{Nasdaq} & \textbf{Crude-oil} & \textbf{Gold} & \textbf{US-dollar} \\ [0.5 ex]
		\hline
		\textbf{Nasdaq}    &-0.3 &	0.02 &	0.05& -0.023 \\\hline
		\textbf{Crude-oil} &-0.00027 & -0.011 & -0.358 &	0.000372  \\\hline
		\textbf{Gold}      &-0.017 & -0.34 & -0.0078 & -0.026  \\\hline
		\textbf{US-dollar} & 0.0037& -0.021& 0.0145& -0.346 \\[1ex]
		\hline
	\end{tabular}
	\caption{Pairwise interactions using drift term of $N$-dimensional Kramers-Moyal coefficient results.}
	\label {table:3}
\end{table}

\begin{figure}[h!]
	\centering
	\includegraphics [width=0.4 \columnwidth] {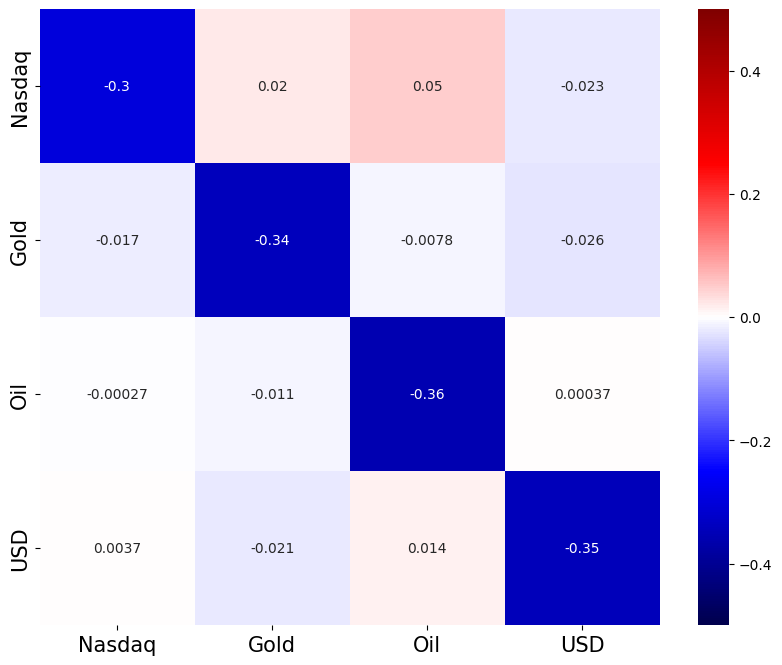}
	\includegraphics [width=0.4 \columnwidth] {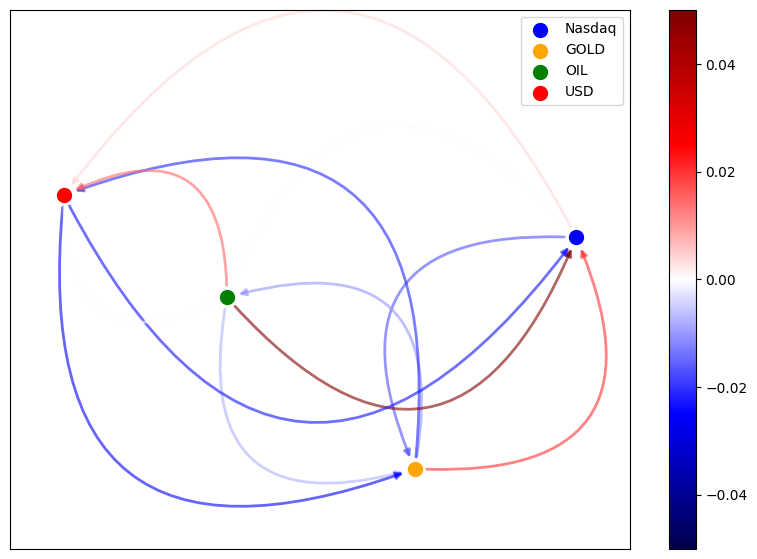}
	\caption{\scriptsize{Linear-Pairwise interactions using drift term of $N$-dimensional Kramers-Moyal  heatmap and symmetric network for crude oil, US dollar, gold and Nasdaq.}}
	\label{fig:4}
\end{figure}
\begin{figure}[h!]
	\centering
	\includegraphics [width=0.9 \columnwidth] {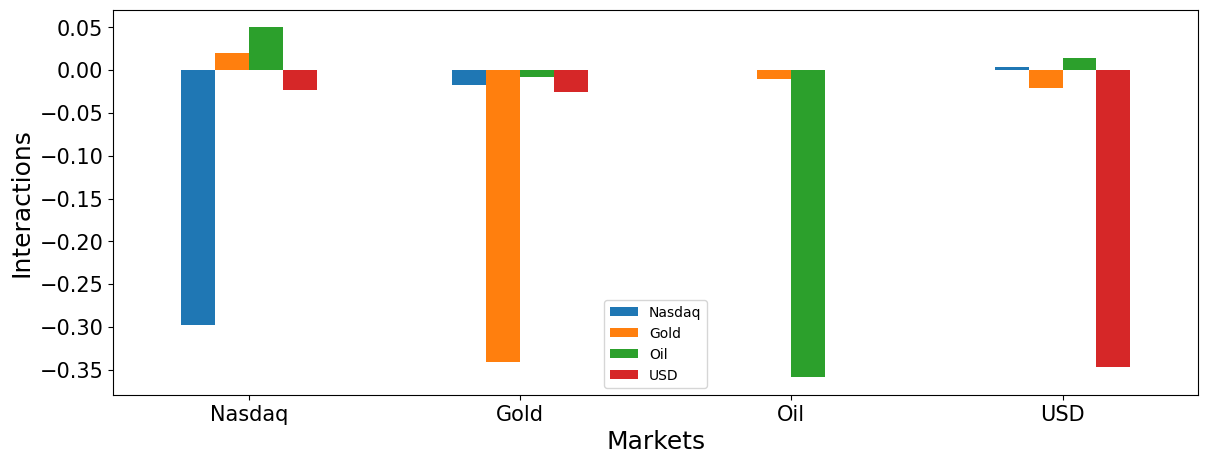}
	\caption{\scriptsize{Linear-Pairwise interactions using drift term of $N$-dimensional Kramers-Moyal barplot for crude oil, US dollar, gold and Nasdaq.}}
	\label{fig:5}
\end{figure}

Utilizing drift term of Krammers-Moyal expansion, one may also consider and analyze the evolution of market in terms of time. Fig. \eqref{fig:6} particularly indicates the modifications in ten continuous sub-intervals of time, including those of Covid-19 pandemic and Ukraine crisis. As it is clearly seen, there are relatively higher amounts of drift coefficients yield in the pairwise interactions during the Covid-19 pandemic, addressing the market's behavior in a more complex chaotic manner.

\begin{figure}[h!]
	\centering
	\includegraphics [width=0.9 \columnwidth] {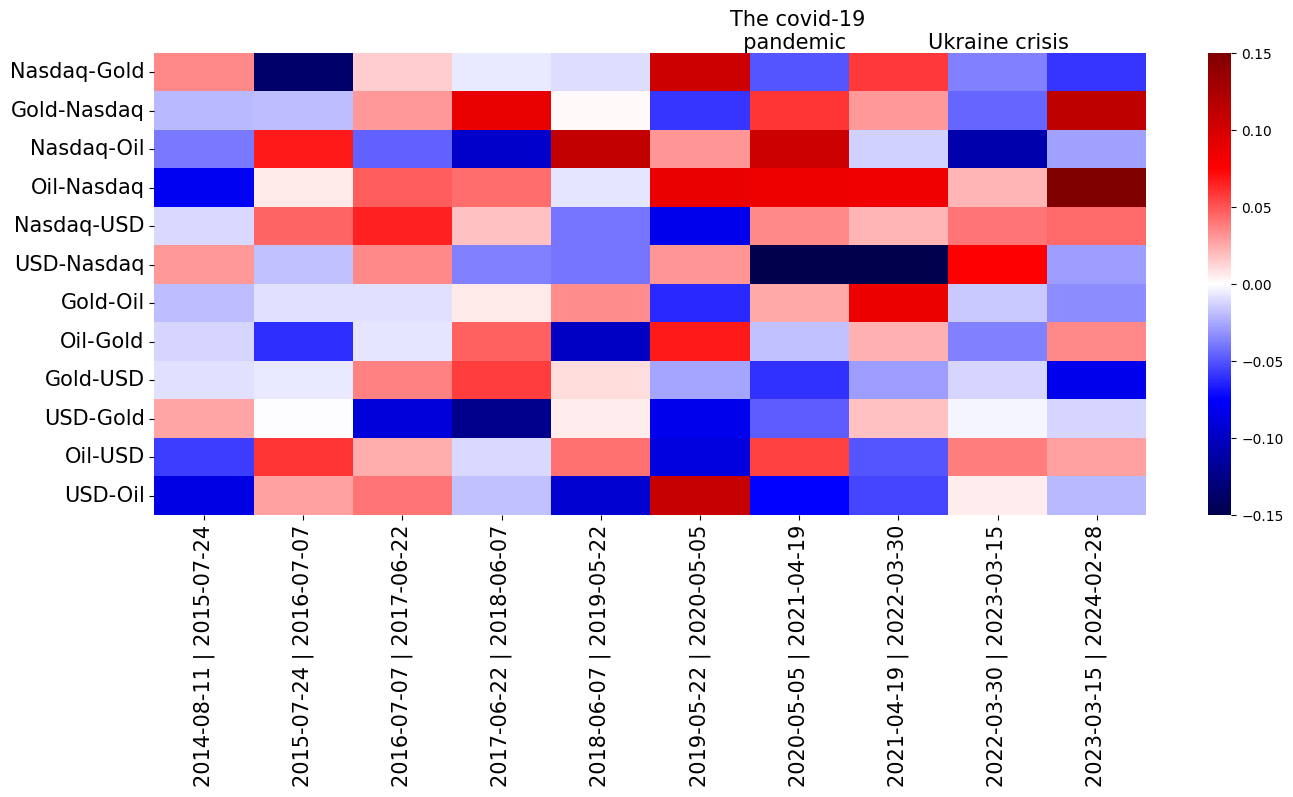}
	\caption{\scriptsize{Linear-Pairwise interactions using drift term of $N$-dimensional Kramers-Moyal evolution heatmap.}}
	\label{fig:6}
\end{figure}

\section{conclusion}
In this research, we attempted different manners to derive pairwise interactions in financial markets. Due to coincident and symmetric nature of mutual entropy and correlation coefficients, these methods gained inadequate information about pairwise behavior. Hence, we alternatively applied transfer entropy and $N$-dimensional Kramers-Moyal expansion to reach a better understanding of dynamics among financial indices. Applying calculated coefficients, we also figured interaction network to better describe the market's system. Nevertheless, the analysis of US dollar index and gold indicated a symmetric interaction dynamic. Finally, time-dependent analysis of indices, especially in Covid-19 pandemic presented a rapid change, confirming a rapid complex behavior during the spread. This study introduces a unified framework that integrates Transfer Entropy (TE) and the N-dimensional Kramers–Moyal (KM) expansion to analyze directional, nonlinear, and stochastic interactions among four major financial indices—the Nasdaq, crude oil, gold, and the U.S. Dollar Index—over a ten-year period encompassing both stable and crisis conditions. By leveraging TE’s ability to capture asymmetric information flows and the KM drift coefficient’s capacity to model deterministic stochastic dynamics, we address key limitations of traditional correlation-based and single-method approaches. Our empirical findings highlight several critical insights. First, we observed significant increases in directional coupling, measured via TE, during crisis periods, particularly the COVID-19 pandemic and the Russia–Ukraine conflict, indicating intensified systemic information flow under stress. Second, KM drift analysis revealed consistent gold–dollar interactions, reinforcing gold’s role as a safe-haven asset. In contrast, oil–equity linkages exhibited structural fragility, weakening during turbulence and rapidly reorganizing thereafter, reflecting regime-dependent dynamics. This dual-method approach demonstrates aspects of market interconnectedness that linear or symmetrical metrics fail to capture. Indeed, it provides both a network-level perspective on information propagation and an asset-specific understanding of stochastic behavior, enabling richer interpretations of market responses to external shocks. Incorporating machine learning techniques for real-time interaction tracking could further enhance predictive capabilities in unstable financial perspective.

\section*{Acknowledgments}
Authors are thankful from Professor M.R. Rahimi Tabar for helping in preparing quantitative methods.

\end{document}